\documentclass{article}
\usepackage[utf8]{inputenc}
\usepackage{jheppub}
\usepackage{amsmath,bm,amsfonts,amssymb,array,calc,amsthm,rotating}
\usepackage{epsfig,psfrag}
\usepackage{pstool}
\usepackage{graphicx}
\usepackage{color}
\usepackage[title]{appendix}

\DeclareMathAlphabet\mathbfcal{OMS}{cmsy}{b}{n}
\definecolor{darkgreen}{RGB}{50,150,0}
\definecolor{purple}{cmyk}{0.5,0.75,0,0}
\definecolor{darkpurple}{RGB}{128,0,128}

\newcommand{\MPl}{M_{\rm Pl}}
\newcommand{\mKK}{m_{\rm KK}}
\newcommand{\mDM}{m_{\rm DM}}
\newcommand{\syncH}{\mathcal{H}}
\newcommand{\vtoday}{v_{\rm today}}
\definecolor{darkgreen}{RGB}{50,150,0}

\newcommand{\mycomment}[1]{}

\title{Dark Dimension and Decaying Dark Matter Gravitons}

\author[a]{Georges Obied,}
\author[b]{Cora Dvorkin,}
\author[c]{Eduardo Gonzalo,}
\author[b]{Cumrun Vafa}

\affiliation[a]{Rudolf Peierls Centre for Theoretical Physics, University of Oxford, Oxford OX1 3PU, United Kingdom}
\affiliation[b]{Department of Physics, Harvard University, Cambridge, MA 02138, USA}
\affiliation[c]{Department of Physics, Lehigh University, Bethlehem, PA, 18018, USA}

\abstract{
We explore the cosmology of the Dark Dimension scenario taking into account perturbations in the linear regime. In the context of the Dark Dimension scenario, a natural candidate for dark matter in our universe is the excitations of a tower of massive spin-2 KK gravitons. These dark gravitons are produced in the early universe and decay to lighter KK gravitons during the course of cosmological evolution. The decay causes the average dark matter mass to decrease as the universe evolves. In addition, the kinetic energy liberated in each decay leads to a kick velocity for the dark matter particles leading to a suppression of structure formation. Using current CMB ({\it Planck}), BAO and cosmic shear (KiDS-1000) data, we put a bound on the dark matter kick velocity today $v_\mathrm{today} \leq 2.2 \times 10^{-4} c$ at 95\% CL. This leads to rather specific regions of parameter space for the dark dimension scenario. The combination of the experimental bounds from cosmology, astrophysics and table-top experiments lead to the range $l_5\sim 1- 10 \, \mu m$ for the size of the Dark Dimension. The Dark Dimension scenario is found to be remarkably consistent with current observations and provides signatures that are within reach of near-future experiments.
}

\begin{document}
\maketitle

\section{Introduction}

The particle nature of dark matter (DM) is one of the major outstanding problems in particle physics and cosmology. There are a plethora of models that are consistent with observations and well-motivated from a particle physics point of view~\cite{ParticleDataGroup:2020ssz}. However, when viewed from an effective field theory perspective, these models leave many questions unanswered such as those related to the electroweak and/or cosmological hierarchy problems. 

Recently, it has been pointed out~\cite{Montero:2022prj} that the Swampland program (see \cite{Agmon:2022thq} for a review) can offer a new take on the cosmological hierarchy problem, leading to a unification of dark energy and dark matter. In the context of the Swampland program, it is natural for there to be a tower of particles with mass close to the energy scale set by the cosmological constant:
\begin{align}
    m \sim \Lambda^\alpha,
\end{align}
in Planck units and with $\alpha \sim \mathcal{O}(1)$. This is essentially the statement of the (A)dS distance conjecture~\cite{Lust:2019zwm}. 

In~\cite{Montero:2022prj}, it was shown that, when this line of thought is followed to its logical conclusion, one is led to an essentially unique model of our universe, dubbed the Dark Dimension. In more detail, it was argued that theoretical and experimental constraints select $\alpha = 1/4$ and allow for only a single tower of Kaluza-Klein states. This has led to the idea of viewing dark matter as excitations in this tower ~\cite{Gonzalo:2022jac}:  The light matter consists of  the Standard Model fields in addition to a tower of massive KK gravitons. Each particle in the tower is coupled only gravitationally to the Standard Model and to other particles in the tower. In the presence of the KK graviton tower and the (necessary) gravitational couplings, it was also shown in~\cite{Gonzalo:2022jac}, that a population of KK gravitons will be produced in the early universe. leading to a natural candidate for dark matter,  which can constitute all of the dark matter. This led to a concrete unification of dark matter and dark energy. Towers of particles have also been considered previously in the context of dark matter models, motivated by their genericity in string theory `the dynamical dark matter scenario' (see for example~\cite{Dienes:2011ja,Dienes:2011sa}). The Dark Dimension is then a well-motivated model from a theoretical perspective given its consistency with Swampland principles and its potential to provide a new take on the cosmological hierarchy problem and its unification with dark matter. It also has phenomenologically interesting signatures as well as possible alternative scenarios that were investigated in various recent papers~\cite{Anchordoqui:2022svl,Anchordoqui:2023oqm,Anchordoqui:2023tln,Cribiori:2023swd,Blumenhagen:2022zzw}.

The purpose of this work is to explore potentially observable cosmological aspects of the Dark Dimension model. In particular, we focus on studying perturbations in the linear regime within a simplified version of the full model that nonetheless captures the important physics. Remarkably, we will find that the model, although borne out of very general string theoretic principles, is consistent with current observations and provides signatures that are within reach of near-future experiments. 

This paper is organized as follows. In \S\ref{sec:KKgravitonPheno} we review the relevant phenomenology of the dark dimension. In this section we introduce some details of the `kick' velocity of the dark matter and distribution functions that we use in our analysis. In \S\ref{sec:CosmologyResults}, we present our results and discuss the effects that play an important role in the constraint that we get. Then, in \S\ref{sec:ModelParameters}, we discuss the parameters of the dark dimension model in light of the new constraints derived from cosmology as well as other experimental data. Finally, we give some concluding remarks and discuss some open questions in \S\ref{sec:conclusion}.

\section{KK Gravitons as Dark Matter}
\label{sec:KKgravitonPheno}

Let us begin by reviewing the essential features of the Dark Dimension scenario. For more details, we refer the reader to~\cite{Gonzalo:2022jac}. In this scenario, the universe has a tower of massive gravitons with spacing:
\begin{align}
    \mKK \sim \Lambda^{1/4} \sim 10-100\  \mathrm{meV}.
\end{align}
These are the Kaluza-Kelin (KK) modes of the graviton in a mesoscopic dimension of length $\l \lesssim 30 \;\mathrm{\mu m}$. The particles of the Standard Model (SM) live on a brane localized in the extra dimension and their coupling to the graviton KK modes is determined (up to $\mathcal{O}(1)$ numbers) by the 5D equivalence principle. This allows the KK gravitons to be produced from a hot brane in the early universe. In particular, if we start with an empty extra dimension (i.e. KK graviton sector) and a SM brane at temperature $T_i\sim 1\;\mathrm{GeV}$, we produce the correct abundance of KK gravitons to account for all of the observed dark matter. The coupling between each KK graviton and the SM is of gravitational strength. Given this weak coupling, the tower states play an essential role in this production mechanism since it would be difficult to produce a sufficient amount of DM with only gravitational strength couplings to a few particle states without going close to Planckian temperatures. The large number of particles in the tower, allow for the production of enough DM at temperatures much lower than Planckian temperature. 

Generically, the extra dimension need not be homogeneous~\cite{Mohapatra:2003ah} and this inhomogeneity will induce a coupling between the different graviton KK modes. In particular, this coupling will allow heavier KK gravitons to decay to lighter ones, again via gravitational strength interactions. The large number of graviton modes available for this decay enhances this decay width relative to decays to the SM (see~\cite{Gonzalo:2022jac} for more details). Thus, when massive gravitons decay, they predominantly decay to lighter gravitons rather than SM particles\footnote{The decay to SM of course also occurs and has been investigated in~\cite{Law-Smith:2023czn}.}. We parametrize the energy scale of this inhomogeneity by $\delta \cdot m_{KK}$ where $\delta \sim \mathcal{O}(1)$, which determines the violation of KK number conservation. More precisely, a parent particle with mass $m_k$ can decay to two daughter particles with masses $m_j$ and $m_l$ such that $m_k = m_j + m_l + \epsilon$ with $\epsilon \leq \delta \cdot m_\mathrm{KK}$. However, in practice we will simply take $\epsilon \approx \delta \cdot \mKK$. Given our lack of knowledge of the precise spacing of the tower (which will depend on the inhomogeneity), we will in general also allow for non-integer values of $\delta$. Finally, while $\delta$ captures the wavelength of the inhomogeneity of the extra dimension, we introduce another effective parameter $\beta$ that captures the amplitude of these inhomogeneities.  This in particular controls the rate of decay of KK gravitons within the tower.

\begin{figure}[hbtp!]
\centering
	\includegraphics[width=0.7\textwidth]{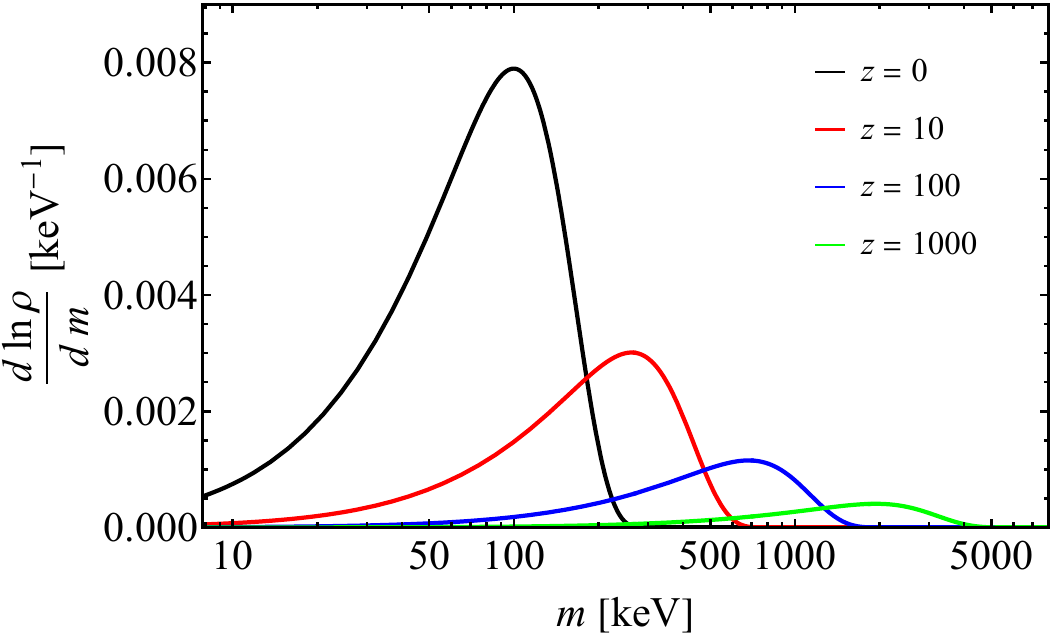}
	\caption{
        The mass distribution of KK gravitons at different times in cosmic history. The horizontal axis should be thought of as a discretum with spacing $\mKK$. The distribution starts at heavier masses and moves to lower masses as the gravitons decay to lighter gravitons. 
	}
	\label{fig:MassSpectrum}
\end{figure}

Most dark gravitons are produced at the initial temperature when $T\sim T_i$ with a mass $m \sim T_i$, and it is a very good approximation to ignore further graviton production at lower temperatures. By then, the most important effects are the decays. Our interest here is in decays within the KK graviton tower. These cause the DM occupation numbers in each mode to vary in the course of cosmological evolution and effectively lead to a time dependent dark matter mass (see Figure~\ref{fig:MassSpectrum}). As the particles decay, the average dark matter mass decreases with time and the distribution creeps to lower masses while keeping its shape. At a given time $t$, the peak of the DM mass distribution is at:
\begin{align}
    \label{eq:DMmass}
    \mDM(t) \approx \frac12 \left(\frac{\MPl^4 \mKK}{\delta^3 \beta^4}\right)^{1/7} \frac{1}{t^{2/7}},
\end{align}
where the parametric dependence can be deduced from the total decay width~\cite{Gonzalo:2022jac} and the prefactor is fixed from numerical simulations.

The intra-tower graviton decays also lead to another important effect: dark matter velocities. When a parent KK graviton decays, it imparts a kick velocity to the daughter particles that can affect cosmological structure formation. Intuitively, these particles can escape gravitational potential wells and lead to less structure in the late-universe. The kick velocity can be easily determined from the kinematics of the problem and using the knowledge that the coupling between KK particles is gravitational. This latter fact implies that the most likely decay products are the most massive ones. In other words, the gravitational nature of the decay means that when a KK graviton of mass $m$ decays to two daughter particles, the most likely outcome is two particles of masses close to $m/2$ (up to the mass violation $\epsilon = \delta \cdot \mKK$ discussed above). For simplicity, we will assume this to be always the case and neglect subdominant effects from asymmetric decays where one of the daughter particles is much heavier than the other. Finally, we will be working in the non-relativistic regime since relativistic dark matter is strongly ruled out and this approximation will be shown to be self-consistent given the experimental constraint derived below.

Under these assumptions, the kick velocity is determined by the mass of the parent particle alone. We consider the decay of a KK graviton of mass $m$ and treat $\epsilon = \delta \cdot \mKK$ as small compared to $m$. This is a very good approximation (as we will see later) since $\epsilon \sim \mKK \sim \mathrm{meV}$, while $m \sim 100\;\mathrm{keV}$. The velocity of the daughter particles is deduced by equating their kinetic energy to the difference in masses, i.e. $\delta\cdot \mKK$. This gives:
\begin{align}
    \label{eq:DMvel}
    v \approx \sqrt{\frac{\delta \mKK}{m_{\text{DM}}}} \propto t^{1/7},
\end{align}

where the time dependence is implied by the time dependence of $\mDM$. We take the velocity as a function of time to be given by Eq.~\eqref{eq:DMvel}. One might worry that we should incorporate the history of kick velocities. The justification for ignoring this history follows from two effects. The first is that velocity redshifts as the universe expands, whereby $v$ decreases as $a^{-1}$. The second is that the kick velocity increases with time as $t^{1/7}$. These two effects, in addition to the fact that we are working in the non-relativistic regime, imply that we can ignore the history of particle kicks and focus only on the kick velocity that a particle receives in the last decay. See also appendix~\ref{app:velDist} for more details.

To sum up, we have three\footnote{There are two other free parameter of the model: (i) $\lambda$,
which controls decay back to SM and (ii) $\gamma$, which determines the mass of the lightest particle in the tower in terms of $\mKK$. The first parameter can be neglected 
for this investigation as the decays to SM particles have little effect on the evolution of perturbations on large scales. In a companion paper~\cite{Law-Smith:2023czn}, we showed that astrophysical bounds are only compatible with natural values of $\lambda \sim O(1)$ if the dark matter mass today is lighter than about $0.3 $ MeV. The second parameter is also irrelevant for cosmology but can have an impact on the interpretation of bounds from fifth force experiments. See \S\ref{sec:ModelParameters} for a more extended discussion of these two additional parameters.} parameters in our model relevant to the analysis in this paper: $\mKK$, $\delta$ and $\beta$.  However, the data we use is only sensitive to the combination that gives the velocity in Eq.~\eqref{eq:DMvel}, and that is the single parameter we implement and constrain in the bulk of this paper. Only then, we will use these constraints along with others to derive bounds on the dark dimension model itself. These will be discussed in \S\ref{sec:ModelParameters}. 

At this point, it is worth emphasizing again that we will only make use of theory and data in the linear regime. As such, our analysis will not capture the implications our model might have on the $\sigma_8$ tension. In fact, in the linear regime, the significance of the $\sigma_8$ tension is much lower (see for example Figure~\ref{fig:shearCls}) and there is not really a puzzle to resolve. That said, dark gravitons cause a suppression in the amount of fluctuations on small scales and go in the right direction to resolve the $\sigma_8$ tension. It would be interesting to study this effect more closely, by running $N$-body simulations that capture the essential phenomenology such as the kick velocities discussed above in the non-linear regime. However, this is beyond the scope of this work. Finally, we mention that the loss in mass of the dark graviton model is reminiscent of that studied in~\cite{Agrawal:2019dlm} and may have implications for the Hubble tension as well.

\subsection{Kick Velocities}

We have already argued that the mean velocity of DM in our model behaves as in Eq.~\eqref{eq:DMvel}. There is also a typical spread in velocities which is of the same order of magnitude as the mean velocity. These two features of our model are important for phenomenology and we capture them by assuming the following background distribution of DM momenta:
\begin{align}
    \label{eq:distributionFunction}
    f_0(p, t)=C \frac{1}{a(t)^3} \frac{1}{\vtoday^3}\left(\frac{t}{t_\mathrm{today}}\right)^{-3 / 7} \exp \left[\frac{-\left(p-p_0(t)\right)^2}{2 \eta p_0(t)^2}\right],
\end{align}
where $p_0 = m \vtoday (t/t_\mathrm{today})^{1/7}$ and we have added an $\mathcal{O}(1)$ constant $\eta$ to parametrize the variance of the distribution in terms of the mean. In all our numerical simulations we will set $\eta = 0.25$, which follows from the behaviour of the velocity distribution at late times (see Appendix~\ref{app:velDist} for more details). The time-dependent normalization of this function is chosen such that the DM energy density redshifts like $a^{-3}$. As shown in~\cite{Gonzalo:2022jac}, this is a very good approximation despite the decay of DM particles and the subsequent loss of the daughter particle kinetic energy as the universe expands. In fact, the fractional energy loss in one Hubble time is:
\begin{align}
    H^{-1} \frac{d \log \rho}{dt} \approx \frac{\delta \mKK}{\mDM} \approx v^2.
\end{align}
We will see later that we need $v \sim 10^{-4}$, implying a negligible loss in DM energy density due to the decays. Said differently, the energy loss is an $\mathcal{O}(v^2)$ effect which is subdominant to the $\mathcal{O}(v)$ velocity effects that we study in this work.

We study perturbations by taking:
\begin{align*}
    f_0 \rightarrow f_0 (1 + \Delta),
\end{align*}
where the Boltzmann equation determines the evolution of $\Delta$. Rather than tracking the evolution of the full distribution function~\eqref{eq:distributionFunction} and its perturbation via a Boltzmann hierarchy, we work in the fluid approximation and deal with moments of $f_0$ and its perturbation. We will truncate the Boltzmann hierarchy at Legendre multipole $l_\mathrm{max} = 2$ (see for example~\cite{Shoji:2010hm}). For a different approach see for example the treatment of~\cite{Lesgourgues:2011rh}. The relevant definitions and equations governing the perturbations are the familiar ones and are reproduced in Appendix~\ref{app:fluidEqs} for completeness.

\section{Analysis with Cosmological Data}
\label{sec:CosmologyResults}

The dark matter we consider has a few novel features compared to particle physics models often considered in the literature (see e.g.~\cite{ParticleDataGroup:2022pth}). First, our DM starts with small velocity which then increases over time. This is an unusual behaviour from the point of view of cosmology, where velocities and temperatures decrease with the expansion of the universe. Moreover, since all the dark gravitons participate in decays, it is all of the dark matter that acquires larger velocities with time rather than just a fraction. In addition, our model does not have a dark radiation component unlike models of decaying dark matter which often include dark radiation as well. Finally, on a more detailed level, the one-seventh power in Eq.~\eqref{eq:DMvel} may allow future experiments to distinguish this model from other models of decaying dark matter (see below for more on this). 

The features mentioned above play an important role in the phenomenology. The main effect is on the evolution of density perturbations (and the related gravitational potentials). In general, one has to do a full $N$-body simulation to capture the non-linear behaviour of a model. In this study we will only focus on the linear regime (see also \S\ref{sec:weakLensing} for more details). In this regime, we will study effects on the cosmic microwave background (CMB) and large-scale structure of the universe. That the CMB is affected is easy to see through the line-of-sight formalism where the gravitational potentials enter explicitly in the calculation of the temperature anisotropies. The effect on the large-scale structure that we study can be traced back to a suppression in the amount of structure on small (but still linear) scales. We will probe these effects indirectly via the changes they induce in the CMB and weak lensing. We rely on the {\it Planck} 2018 for measurements of the CMB and KiDS-1000 for weak lensing (see also \S\ref{sec:dataandcode} below). 

\begin{figure}[hbtp!]
\centering
	\includegraphics[width=0.7\textwidth]{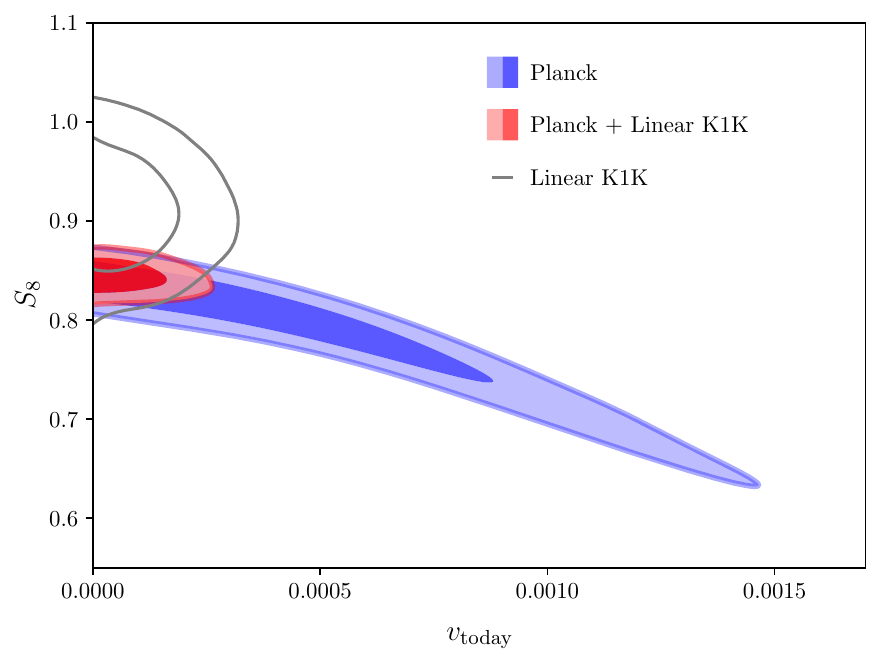}
	\caption{
        We show the 68\% and 95\% CL constraints in the $(\vtoday, S_8)$ plane when using various datasets. The Planck dataset includes high-$\ell$, low-$\ell$ temperature and polarization data as well as lensing. The Linear K1K data are described in \S\ref{sec:weakLensing}. In addition, all data combinations include the BAO data described in \S\ref{sec:dataandcode}. The contours show how K1K and {\it Planck} data provide complementary constraints on $\vtoday$ and $S_8$. 
	}
	\label{fig:vtoday_vs_S8}
\end{figure}

Using {\it Planck} data, we get a robust constraint on $\vtoday$:
\begin{align}
    \label{eq:vConstraintCMB}
    \vtoday < 1.2 \times 10^{-3} \qquad \text{(95\% CL, \it Planck)}.
\end{align}
These constraints are derived from two effects that alter the CMB. First the intrinsic unlensed CMB is suppressed at high multipole due to the presence of the graviton velocity which suppresses fluctuations. The other important effect is due to alterations of the lensing potential which relies on the intermediate structure between us and the surface of last scattering. In addition, as can be seen from Figure~\ref{fig:vtoday_vs_S8}, there's a negative correlation between $S_8$ and $\vtoday$ as expected. This is due to the intuitive fact that larger velocities would lead to less structure. The {\it Planck} measurement of $S_8$ (and other cosmological parameters) has smaller uncertainty and, when combined with weak lensing data, {\it Planck} fixes these remaining cosmological parameters. 

Let us now briefly discuss the constraints when adding weak lensing measurements. As mentioned previously, we will only make use of linear scales. Since this cut off between linear and non-linear scales cannot be defined in a precise way, the constraints on $\vtoday$ depend on where we make this cut. More details on this is provided in \S\ref{sec:weakLensing}. That said, the 95\% CL constraints are:
\begin{align}
    \label{eq:vConstraintLensing}
    \vtoday \leq 10^{-4} \times \left\{
    \begin{array}{lr}
        2.2,  & \qquad \Delta_\mathrm{NL} = 0.5,\; k_{\mathrm{NL},0} = 0.09\;\mathrm{Mpc}^{-1}  \\
        1.1, & \qquad \Delta_\mathrm{NL} = 1.0,\; k_{\mathrm{NL},0} = 0.17\;\mathrm{Mpc}^{-1}. 
    \end{array}
    \right. 
\end{align}
where the definition of $\Delta_\mathrm{NL}$ is given in \S\ref{sec:weakLensing} and the subscript $0$ indicates that this is the largest $k$ in the linear regime today. The mode numbers shown for $\Delta_\mathrm{NL} = \{0.5,1.0\}$ are consistent with those found in~\cite{Foreman:2015lca}. It is easy to understand where this constraint comes from intuitively. The graviton kick velocities allows them to escape potential wells that would otherwise cause them to collapse into bound structures. This suppresses the amount of structure on small (but still linear) scales and alters the lensing (convergence) power spectrum that the data measures. A large change that comes with a large value of the velocity parameter $\vtoday$ is disfavored by the data. All in all, and depending on the cut between linear and non-linear scales, the two datasets\footnote{One might worry about combining the two datasets given the known $\sigma_8$ tension. In this work, we restrict our attention to linear scales where the two datasets are consistent at $\sim 1 \sigma$ even under $\Lambda$CDM.} provide complimentary constraints in the $S_8$ and $\vtoday$ plane as shown in Figure~\ref{fig:vtoday_vs_S8}. While we quote the bound for two values of $\Delta_{\rm NL}$ here, we will use the conservative constraint, with $\Delta_{\rm NL} = 0.5$ in the rest of the paper. 

Finally, we comment briefly on related constraints from non-linear scales that have been discussed in the literature~\cite{Wang:2014ina,DES:2022doi}. These constraints come from comparing the known abundance of low-mass subhalos with results from $N$-body simulations that explore galactic substructure. In these decaying dark matter models, one considers a long-lived dark matter candidate. The lifetime of such a particle is taken to be close to the age of the universe $\tau \sim t_H$ and thus early structure formation proceeds in the same way as $\Lambda$CDM. However, once dark matter particles begin to decay (typically at low redshifts because of the chosen lifetime) the evolution can deviate from CDM. This deviation depends on the recoil/kick velocity. Intuitively, if the kick velocity is large enough then it can strip mass off halos and subhalos and make the latter especially susceptible to tidal disruption events. As such, fewer subhalos survive to the present day and one would observe a reduced abundance. The reported bounds~\cite{DES:2022doi} depend in general on the recoil/kick velocities (for example for kick velocities of 20 km/s (40 km/s), a lifetime longer than 18 Gyr (29 Gyr) is required for the dark matter candidate to not be ruled out). However, for recoil velocities larger than the maximum circular velocity of the subhalos, the bound becomes independent of the kick velocity and depends only on the dark matter lifetime. For kick velocities this large, all decays lead to the daughter particle being ejected from the dark matter halo and all that matters is the number of decays which is set by the lifetime. In this case, lifetimes shorter than $\sim$ 40 Gyrs are ruled out~\cite{DES:2022doi}.

Our model is slightly different from models of a single decaying dark particle since our dark matter decays throughout cosmic history rather than just at low-redshift. In addition, as we previously argued, the kick velocity is a function of time as is shown in Eq.~\eqref{eq:DMvel} unlike the constant kick velocities often studied in the literature. However, despite these difference one may hope that the bounds from the literature can be applied to our model conservatively although it would be impossible to draw quantitative conclusions without further analysis. 

Given the discussion above, we check that our model is not ruled out by observations of subhalo abundances. The natural parameter range for our model has a kick velocity that is of the same order as the bounds in Eqs.~\eqref{eq:vConstraintCMB} and~\eqref{eq:vConstraintLensing}, which we can take to be $\vtoday \sim 10^{-4} \sim 30\;\mathrm{km s^{-1}}$. These velocities are similar to the recoil velocities studied in~\cite{DES:2022doi}. As such, we are in the regime where only the lifetime of the dark gravitons matters (as explained above) and the lifetime bound $\tau \gtrsim 40\;\mathrm{Gyr}$ is a constraint on our model. We now show that our model automatically satisfies this bound. 

The dark graviton decay lifetime\footnote{In fact, in our model, we only have a notion of average lifetime which is  parametrically of order the Hubble time but differs by important $\mathcal{O}(1)$ factors such as $\beta$ and $\delta$. } is given by~\cite{Gonzalo:2022jac}:
\begin{align}
    \tau = \frac{\MPl^2 \mKK^{1/2}}{\beta^2 \delta^{3/2} \mDM^{7/2}} = 2^{7/2} t_\mathrm{age} \approx 10 t_\mathrm{age}
\end{align}
where we have substituted the expression for $\mDM$ in terms of $\delta, \beta$ and $\mKK$ from Eq.~\eqref{eq:DMmass} and $t_\mathrm{age} = 13.7\;\mathrm{Gyr}$ is the age of the universe today. We see that the fact that $\mDM$ is slightly lower than the value obtained by equating the decay rate to the Hubble rate (by a factor of 2 in this case) is the reason our model automatically satisfies this bound. As mentioned previously the factor of 2 is obtained from numerical simulations for $\mathcal{O}(1)$ parameters of the model. That said, the following intuitive reasoning indicates that the suppression by a factor larger than one is generic in our model. Consider the situation where $t \approx H^{-1}$, i.e. a Hubble time has passed. At this time, particles with lifetime equal to a Hubble time have already decayed. As particles decay mostly to daughter particles with half the mass (modulo weak dependence on model parameters), the remaining particles after a Hubble time have a mass equal to half the mass of the particle whose lifetime is a Hubble time. As such, the expectation is that the DM has a mass suppressed from the na\"{i}ve estimate by a factor close to two. Simulations show that this factor can depend very weakly on the parameters $\beta$ and $\delta$ (for example it goes from 2 to 3 when ($\beta,\delta$) goes from (3, 1) to (0.4, 10)). As this dependence is very weak, we do not attempt to quantify it further but simply remark that our model will automatically satisfy the lifetime bound if $\mDM$ is suppressed relative to the naive parametric dependence by a factor larger than 1.4.

\subsection{Effect of the Kick Velocity on Cosmology}
\label{sec:KickVelocityEffects}

In this section we will discuss the effects of including a kick velocity of the form~(\ref{eq:DMvel}) on the evolution of cosmic perturbations. We will be mainly interested in small scales that are nonetheless in the linear regime. As usual, the larger scales evolve just as in $\Lambda$CDM (cf. for example the case of warm dark matter~\cite{Avila-Reese:2000nqd,Bode:2000gq} where only small scales differ from the $\Lambda$CDM predictions). The inclusion of the kick velocity will wash out inhomogeneities on small scales, and in this section we discuss the dynamics of this process. We work in synchronous gauge but the translation to any other gauge should be straightforward. 

We start with the equations governing the evolution of dark graviton perturbations (in synchronous gauge):
\begin{align}
\label{eq:deltaf}
    \dot{\delta}_f &= -(1+w)\left(\theta_f + \frac{\dot{h}}{2}\right) - 3 \syncH (c_a^2 - w) \delta_f\\
\label{eq:thetaf}
    \dot{\theta}_f &= -\syncH(1-3c_a^2) \theta_f + \frac{k^2 c_a^2}{1+w}\delta_f,
\end{align}
where the overdot denotes a derivative with respect to conformal time $\tau$ and the subscript $f$ stands for `fluid' as we are treating the dark gravitons in the fluid approximation. In writing the above equations, we have dropped terms proportional to the shear since they only play a subdominant role in the following discussion. That said, these terms can be found in Appendix~\ref{app:fluidEqs} and have been taken into account in the full analysis. The expressions for the equation of state $w$ and the sound speed $c_a^2$ (which is obtained following reference~\cite{Lesgourgues:2011rh}) are given by:
\begin{align}
    \label{eq:wfld}
    w &= \frac{p_0}{\rho_0} \vtoday^2 \left(\frac{t}{t_\mathrm{today}}\right)^{2/7} \\
    \label{eq:soundspeedfld}
    c_a^2 &= \frac{w}{3(1+w)}\left(5 - \frac{\mathfrak{p}_0}{p_0} \vtoday^2 \left(\frac{t}{t_\mathrm{today}}\right)^{2/7}\right),
\end{align}
where $p_0$, $\rho_0$ and $\mathfrak{p}_0$ are the constant pressure, energy density and pseudo-pressure, respectively, and are defined in Appendix~\ref{app:fluidEqs}. In particular, the above equation shows that for small $\vtoday$, we have $c_a^2 \approx 5w/3$ and the adiabatic sound speed $c_a^2$ and equation of state $w$ are of the same order. The cold dark matter (CDM) limit of the above perturbation equations is obtained by taking $\vtoday \rightarrow 0$ which implies $w, c_a^2 \rightarrow 0$. In addition to the above two equations, we need equations for the metric perturbations $\dot{h}$ and all other perturbations that source $\dot{h}$. These are standard and can be found in many treatments of cosmological perturbation theory so we omit a detailed discussion and simply write the schematic form of the $\dot{h}$ equation:
\begin{align}
    \label{eq:hdotEq}
    \frac{1}{a^3} \frac{d}{d\tau}\left(a \frac{\dot{h}}{2}\right) \propto \sum_i\rho_i \delta_i,
\end{align}
where the sum is over all the species present in the universe. 

We now consider the evolution of the dark graviton perturbations $\delta_f$ and $\theta_f$ for a particular comoving wavenumber $k$. The wavenumbers of interest are superhorizon modes deep in the radiation era and enter the horizon in the radiation era. These modes start with the same adiabatic initial conditions as cold dark matter:
\begin{align}
    \delta_{f}^{(i)} = \frac34 \delta_{\gamma}^{(i)} \qquad ; \qquad \theta_{f}^{(i)} \approx 0,
\end{align}
where $\delta_{\gamma}^{(i)}$ is the initial condition for the photon density perturbation and the vanishing initial velocity perturbation is the usual one imposed on CDM in synchronous gauge. The choice of CDM initial conditions is justified in our case because the kick velocity vanishes in the limit $t \rightarrow 0$, so that our graviton dark matter reduces to CDM at very early times. This is also apparent in Eqs.~\eqref{eq:deltaf} and~\eqref{eq:thetaf}, which also reduce to the CDM evolution equations at early times. Since we have the same initial conditions and evolution equations as CDM at early times, the density and velocity perturbations of dark gravitons evolve just like CDM until other non-CDM terms in Eqs.~\eqref{eq:deltaf} and~\eqref{eq:thetaf} become important.

In particular, deep in the radiation era, the gravitational potentials are set by the radiation perturbations (photons and neutrinos) and the dark graviton perturbations simply evolve in these potential wells. While they are still superhorizon, the density perturbations grow as $\delta_f \sim a^2$, just like CDM. However, unlike CDM, the velocity perturbation $\theta_f$, is small but non-zero\footnote{See also appendix~\ref{app:velDist} for a discussion of this velocity perturbation in the context of the synchronous gauge.}. While $\theta_f$ vanishes initially, the second term of Eq.~\eqref{eq:thetaf} is non-zero and causes the $\theta_f$ solution to deviate from its CDM counterpart. The two terms are at the same order and $\theta_f$ is given by the parametric expression:
\begin{align}
    \label{eq:velgrowth}
    \theta_f \sim \frac{k^2 w}{\syncH} \delta_f.
\end{align}
This value is still suppressed (by the smallness of $w$) compared to $\delta_f$ and unimportant at this stage. Intuitively, for superhorizon modes, the pressure is not large enough to affect the evolution of the fluid. That said, as $\delta_f$ grows, the velocity perturbation also grows and this will eventually alter the evolution of this wavenumber so that it deviates from the CDM solution. For the values of $\vtoday$ and wavenumbers relevant to our analysis, this happens in the matter era. 

Let us then continue following our perturbation mode through horizon crossing and into the matter era. A particular wavenumber becomes subhorizon when $\syncH < k$. This happens in the radiation era for modes and parameters of interest to us, as we mentioned previously. In this case, the evolution of $\delta_f$ no longer follows $a^2$ and becomes logarithmic instead, going like $\ln a$. This continues until the universe transitions into matter domination, where $\delta_f$ starts growing like a power law $\delta_f \sim a$ again. Throughout this time, the velocity perturbation is growing as is given by Eq.~\eqref{eq:velgrowth}. This behaviour continues until $\theta_f$ is large enough to compete with the gravitational potential $\dot{h}$ term in Eq.~\eqref{eq:deltaf}. At this point, the pressure (in this mode) can counteract the effect of gravitational collapse and the mode stops growing and starts oscillating instead. As such, this mode ends up today with a lower amplitude than it would have had in the CDM case, i.e. the linear matter power spectrum $P(k)$ is suppressed. Smaller wavelength modes deviate from the CDM behaviour earlier and thus end up with more suppressed fluctuations compared to longer wavelengths.

In the following sections, we will be referring to the Newtonian gauge gravitational potentials, so we briefly review the physics of the suppression of fluctuations in this gauge. Recall first, the Newtonian gauge metric perturbations $\Phi$ and $\Psi$ that appear in the 00 and $ii$ components. We will work in the limit where we ignore the shear implying, as usual, that $\Phi = \Psi$. The analog of equation~\eqref{eq:hdotEq} is now the algebraic Poisson equation that determines the gravitational potential $\Phi$ directly in terms of the matter perturbation variables $\delta_i$ and $\theta_i$. As such, we will focus our discussion on the matter perturbations only. In Newtonian gauge, these are governed by the following equations:
\begin{align}
    \label{eq:deltafNewt}
    \dot{\delta}_f &= -(1+w)\left(\theta_f -3\dot{\Phi}\right) - 3 \syncH (c_a^2 - w) \delta_f\\
\label{eq:thetafNewt}
    \dot{\theta}_f &= -\syncH(1-3c_a^2) \theta_f + \frac{k^2 c_a^2}{1+w}\delta_f + k^2\Phi,
\end{align}
where we are ignoring shear contributions as in the previous discussion in synchronous gauge. Like before, we are interested in modes that enter the horizon during the radiation era. As in the case of the synchronous gauge above, the evolution of these modes splits into two regimes (see for example~\cite{Weinberg:2008zzc}). In the first regime, we focus on horizon entry in a radiation dominated universe. This is followed by a second regime which tracks sub-horizon evolution through matter-radiation equality. 

We start by reviewing the evolution of these modes in the CDM limit (obtained by taking $w, c_a^2 \rightarrow 0$) before moving on to the discussion of dark gravitons. In the standard $\Lambda$CDM model, the CDM density perturbation remains constant for superhorizon modes in the radiation era. After horizon entry in the radiation era, CDM perturbations grow logarithmically (proportional to $\ln a$) and then linearly (proportional to $a$) when the universe becomes matter dominated. Similar to the matter perturbation, the gravitational potential $\Phi$ also remains constant for superhorizon modes in the radiation era. It then decays rapidly in the radiation era (proportional to $a^{-2}$) when it becomes subhorizon. Once the universe becomes matter dominated, the amplitude of this $\Phi$ mode remains constant (until the dark energy epoch, which is beyond our interest here).

Let us contrast the above behaviour to that of our model where $w, c_a^2 \neq 0$. In particular, the $\theta_f$ equation~\eqref{eq:thetafNewt} contains a term proportional to the density perturbation which grows with time. This term increases for two reasons: first because the $\delta_f$ perturbation is growing and, second, because the kick velocities cause the sound speed to increase. As such, there comes a time (usually during the matter era for the modes of interest) where this term can interfere with the standard evolution of $\theta_f$ due to Hubble expansion and the gravitational potential. When this happens, it causes the $\theta_f$ and, in turn $\delta_f$, perturbations to oscillate instead of grow. As such these density perturbations end up with a smaller amplitude today (compared to their CDM counterparts) and the matter power spectrum is suppressed on these scales. Relatedly, the gravitational potentials, which are determined algebraically in terms of $\delta_i$ and $\theta_i$, are also suppressed. 

In summary, the growth of the velocity perturbation for graviton dark matter leads to a suppression in fluctuations on small scales. The details of this mechanism are quantitatively different from other models relying for example on the $t^{1/7}$ factor in Eq.~\eqref{eq:DMvel}. This will distinguish signals of dark gravitons from other mechanisms that also suppress the power spectrum (for example warm dark matter) allowing future experiments to potentially differentiate the Dark Dimension scenario from other models. 

The cosmological effects we discuss in this paper can all be traced back to this suppression in density (and relatedly gravitational potential) perturbations on small scales compared to their $\Lambda$CDM counterparts. Ultimately, this leads to a suppression in the matter power spectrum and this may help alleviate tensions between large-scale structure and CMB data within the $\Lambda$CDM paradigm. That said, we do find evidence of this in the linear regime, and a more in-depth analysis into the non-linear regime is required to shed light on these tensions. We now turn to a brief discussion of these effects.

\subsection{Datasets}
\label{sec:dataandcode}

In order to compare our model to cosmological data, we modify the publicly available Boltzmann solver \texttt{CLASS}~\cite{Blas:2011rf} and interface it with the sampler \texttt{MontePython}~\cite{Audren:2012wb,Brinckmann:2018cvx}. The modification adds one additional parameter $\vtoday$ that we vary alongside the usual $\Lambda$CDM parameters $\{A_s, n_s, \theta_*, \Omega_bh^2, \Omega_ch^2, \tau_\mathrm{reio}\}$. We use the following datasets in various combinations:
\begin{itemize}
    \item {\it Planck} 2018: we use the DR3 version of publicly available likelihood code released by the {\it Planck} collaboration. In particular, we use high- and low-$\ell$ temperature and polarization data as well as the {\it Planck} lensing measurements~\cite{Planck:2018vyg,Planck:2018lbu,Planck:2019nip}.
    \item BAO: we use the combined galaxy and Lyman-$\alpha$ BAO likelihood released in~\cite{Cuceu:2019for}. This likelihood uses data from eBOSS DR14 galaxy BAO measurements and auto- and cross- correlation of Lyman-$\alpha$ absorption with quasars~\cite{deSainteAgathe:2019voe,Blomqvist:2019rah}. 
    \item KiDS-1000 weak lensing measurements: we use the cosmic shear measurements of KiDS-1000~\cite{Kuijken:2019gsa} in the form of $E$- and $B$-mode band powers. The bandpower spectra are implemented in a \texttt{MontePython} likelihood~\cite{KiDS:2020suj} and can be used for MCMC simulations in a KiDS Cosmology Analysis Pipeline (KCAP) environment~\cite{Joachimi:2020abi}. The power spectra are obtained by correlating cosmic shear measurements in $5$ tomographic redshift bins for a total of $15$ independent cross- or auto-correlations. We modify the publicly available likelihood to make use of only the linear scales. More on this is discussed in \S\ref{sec:weakLensing}. 
\end{itemize}

We perform a Markov Chain Monte Carlo likelihood analysis, where the sampling is done using the Metropolis-Hastings algorithm, except when deriving constraints from the KiDS-1000 likelihood only where we use nested sampling~\cite{Feroz:2007kg,Feroz:2008xx,Feroz:2013hea}. Finally, when running chains using only the KiDS-1000 data, we impose the priors listed in Table~\ref{tab:priors} that match those used in~\cite{KiDS:2020suj}, for example, as well as other KiDS analyses. In addition, we fix the $\Lambda$CDM parameters $n_s$ and $\tau$ to their {\it Planck} measured values as KiDS-1000 data does not constrain them.

\begin{table}[h!]
\centering
\begin{tabular}{ |c|c| } 
 \hline
Parameter & Priors \\
 \hline
$S_8$ & [0.1, 1.3] \\
$\omega_b$ & [0.019, 0.026] \\
$\Omega_f$ & [0.1, 0.5] \\
$h$ & [0.64, 0.82] \\
$n_s$ & 0.966\\
$\tau_\mathrm{reio}$ & 0.0543 \\
 \hline
\end{tabular}
\caption{Table of priors imposed on the varying parameters when only KiDS-1000 and BAO data are used. The parameter $\Omega_f$ is the density parameter for dark gravitons which play the role of dark matter in our model. The last two parameters ($n_s$, $\tau$) are fixed to their {\it Planck} best-fit when only KiDS-1000 and BAO data are used since they are otherwise poorly constrained.}
\label{tab:priors}
\end{table}

\subsection{Effects on the CMB}

The CMB can be potentially altered in two ways by the dark graviton kick velocities. First, there is a possible change to the intrinsic unlensed CMB. Neglecting large scale effects (such as the late ISW effect), this will happen on scales that begin to deviate from their $\Lambda$CDM behaviour at or before recombination ($z\approx 1100$). This happens at very small scales on the order of $10 - 20\;\mathrm{Mpc}^{-1}$ (recall that larger scales deviate from $\Lambda$CDM at later times) and in fact much smaller than those accessible to current observations. 

The second important effect is gravitational lensing. Here, we can divide the story into two parts: the lensing power spectrum itself and the lensing of the CMB. The lensing potential quantifies the deviation of light geodesics as they approach an observer can be calculated by a line-of-sight integral:
\begin{align}
    \label{eq:lensingPotential}
    \phi(\hat{\bf n}) = -2 \int_0^{\chi_\mathrm{CMB}} d\chi' \frac{\chi_\mathrm{CMB} - \chi'}{\chi_\mathrm{CMB} \chi'} \Phi(\chi \hat{\bf n}, \tau = \tau_\mathrm{today} - \chi'),
\end{align}
where $\tau$ is conformal time, $\chi$ is the comoving angular diameter distance and $\Phi$ is the Newtonian potential. It is clear that the lensing potential will incur large changes when the Newtonian potential $\Phi$ varies. In particular, in our case, where the Newtonian potential in some modes decays at late times, we expect to observe a smaller value of $\phi$ and its variance on those scales. A large suppression in the lensing power spectrum is disfavored by the {\it Planck} lensing likelihood and this is partially responsible for our model constraints.

The lensing potential defined in Eq.~\eqref{eq:lensingPotential} is also responsible for lensing of the CMB (see for example~\cite{Okamoto:2003zw,Lewis:2006fu}). Schematically, the observed lensed temperature anisotropy $\Theta(\hat{\bf n}) \equiv \Delta T(\hat{\bf n})/T$ is related to its unlensed couterpart $\tilde{\Theta}(\hat{\bf n})$ by:
\begin{align}
    \Theta(\hat{\bf n}) = \tilde{\Theta}(\hat{\bf n} + \nabla\phi) \approx \tilde\Theta(\hat{\bf n}) + \nabla\phi \cdot \nabla \tilde\Theta(\hat{\bf n})
\end{align}
in the limit of weak lensing. A smaller lensing potential leads to less distortion of the CMB and sharper peaks in the power spectra. For $\vtoday\sim 10^{-3}$, this effect contributes an $\mathcal{O}(1 \%)$ in the power spectrum.

Finally, we discuss how the above effects change the goodness of fit by appealing to the {\it Planck} 2018 likelihoods. To do that, we compare two models, one with $\vtoday \ll 10^{-3}$ (the $\Lambda$CDM limit) and the other with $\vtoday \approx 10^{-3}$. This latter one is at the boundary of the 95\% CL region shown in Eq.~\eqref{eq:vConstraintCMB}. We use the full TTTEEE likelihood as well as versions with the $\ell$ range restricted to $\ell < 800$ and $\ell < 1600$ to understand what region of the CMB drives these constraints. The results are presented in Table~\ref{tab:CMBlogLikelihoods}. We see that a change in $\vtoday$ (while keeping all other parameters fixed) affects the CMB on all scales. In addition, the fit to the reconstructed lensing power spectrum also deteriorates as shown by the {\it Planck} lensing likelihood. Altering the cosmology cannot improve the CMB fit by much and the largest improvement is to the lensing power spectrum. Nuisance parameters on the other hand improve the CMB fit for the most part. Altogether, the overall fit remains slightly worse than that with lower $\vtoday$, resulting in an upper bound on the kick velocity shown in Eq.~\eqref{eq:vConstraintCMB}.

\begin{table}[h!]
\centering\begin{tabular}{ |l|c|c|c|c|c| } 
 \hline
 Likelihood & $\Lambda$CDM & $\vtoday \rightarrow 10^{-3}$ & Cosmology & Nuisance & $\vtoday \approx 10^{-3}$ best-fit  \\
 \hline
{\it Planck} T+P ($\ell < 800$)  & 1109.23  & +0.93 & -0.15 & -0.05 & 1109.95 \\
{\it Planck} T+P ($\ell < 1600$) & 2117.17 & +2.03 & -0.47 & -0.06 & 2118.68 \\
{\it Planck} T+P                 & 2347.16 & +2.80 & -0.40 & -0.55 & 2349.01 \\
{\it Planck} lensing             & 8.74    & +0.93 & -0.44 & -0.01 & 9.22 \\
 \hline
\end{tabular}
\caption{Breakdown of $-2\ln\mathcal{L}$ contributions in going from one best-fit model to another along a path where we change one set of parameters at a time. The two best-fit models are: (i) a $\vtoday \ll 10^{-3}$ best-fit model (the $\Lambda$CDM limit) and (ii) the best-fit model with a fixed $\vtoday \approx 10^{-3}$. Fitting is done to full {\it Planck} data in addition to the BAO dataset discussed in \S\ref{sec:dataandcode}. The table is best read from left to right with changing parameters indicated in the headers and the differences in $-2\ln\mathcal{L}$ in the table. The first and last column show the value of $-2\ln\mathcal{L}$. The benchmark value $\vtoday \approx 10^{-3}$ is chosen since it lies within the 95\% CL constraint using {\it Planck} data (see Figure~\ref{fig:vtoday_vs_S8}).}
\label{tab:CMBlogLikelihoods}
\end{table}

\subsection{Effects on Weak Lensing}
\label{sec:weakLensing}

The observed image of a galaxy can be gravitationally lensed by intervening cosmic structure along the line of sight to the galaxy. This can cause two types of distortions to the original image (for a review see~\cite{Kilbinger:2014cea}). The first is an isotropic stretching of the image, which is quantified by the \emph{convergence}. The second is an anisotropic stretching altering the shapes of galaxies and is quantified by the \emph{shear}. This decomposition of the shear field is exactly analogous to the decomposition of the CMB temperature anisotropies into curl-free $E$-modes and divergence-free $B$-modes. 

In the case of weak lensing, the $B$-mode auto-correlation spectrum is expected to vanish in the absence of systematics. As such, we focus our discussion on the $E$-mode (convergence) power spectrum, although we use the $E$- and $B$-mode power spectra supplied by KiDS-1000 collaboration~\cite{Kuijken:2019gsa,KiDS:2020suj} in our analysis. In the Limber approximation, the convergence power spectrum can be expressed in terms of a line-of-sight integral over the matter power spectrum multiplied by two weight functions $q_\mu(\chi)$ (see also~\cite{Kohlinger:2017sxk,Kilbinger:2017lvu} for more details):
\begin{align}
    \label{eq:convergenceCls}
    C_{\mu\nu}(\ell) = \int_0^{\chi_H} d\chi \frac{q_\mu(\chi) q_\nu(\chi)}{\chi^2} P_m\left(k = \frac{\ell + 0.5}{\chi}; \chi\right),
\end{align}
where $\chi_H$ is the comoving distance to the horizon and $P_m$ is the matter power spectrum. The window functions quantify the lensing efficiency and contain information about the galaxy redshift distribution. They are given by
\begin{align}
    q_\mu(\chi) = \frac{3\Omega_m H_0^2}{2}\frac{\chi}{a(\chi)} \int_\chi^{\chi_H} d\chi' n_\mu(\chi') \frac{\chi' - \chi}{\chi'},
\end{align}
where $n_\mu(\chi)$ is a normalized source redshift distribution. The presence of two weight functions is because we are computing a two-point correlation function. The indices $\mu,\nu$ indicate that, depending on the chosen weight functions, one can compute the convergence power spectrum between different source populations. For instance, in the KiDS-1000 analysis that we follow here, the galaxy sample is divided into 5 tomographic redshift bins and one considers the 15 independent convergence correlations between these bins. In this case, there are five choices for each weight function and they quantify the galaxy redshift distribution in each tomographic bin. 

The original KiDS-1000 analysis makes use of non-linear approximations to the matter power spectrum $P_m$ (for example using Halofit~\cite{Peacock:1996ci,Smith:2002dz} or HMCode~\cite{Mead:2015yca,Mead:2020vgs}). However, our model is expected to be different from $\Lambda$CDM on non-linear scales and the prescriptions for deducing the non-linear power spectrum from the linear one (as is done in Halofit or HMCode) need to be modified. We leave this for future work and focus instead on the linear regime approximating $P_m$ by the linear power spectrum $P_m^{\rm lin}$. In addition, we need to ensure that we only use information from linear scales and we do this by restricting the multipole range in the convergence (and shear) power spectra according to the following prescription. 

We consider a comoving scale $k$ at redshift $z$ to be non-linear if it is larger than $k_\mathrm{NL}$, where the latter is given by the following criterion:
\begin{align}
    \frac{k_\mathrm{NL}^3 P_m^{\rm lin}(k_\mathrm{NL}, z)}{2\pi^2} = \Delta_\mathrm{NL},
\end{align}
where $P_m^{\rm lin}(k,z)$ is the linear power spectrum and we take three values for $\Delta_\mathrm{NL} \in \{0.5, 1.0\}$. Doing this allows us to quantify the dependence on $k_\mathrm{NL}$ since the cutoff between linear and non-linear scales is not sharp. We emphasize that $k_\mathrm{NL}$ depends on redshift since the linear power spectrum grows with time. In practice, the above restriction to linear scales means that we only use certain low multipoles for each redshift and discard all the higher ones. When the spectrum in question is a cross-correlation between two redshift bins, then we take the more conservative choice for the multipole range (i.e. only using the data points that are linear at both redshifts). We refer to this restricted data set by the name `Linear K1K' and use it instead of the full KiDS-1000 data when constraining our model. Finally, we note that for $\Delta_\mathrm{NL} \in \{0.5, 1.0\}$, we get $k_{\mathrm{NL}} \in\{0.09, 0.17\}\;\mathrm{Mpc}^{-1}$ at $z=0$, which is roughly when the linear approximation begins to breakdown. 

\begin{figure}[hbtp!]
\centering
	\includegraphics[width=1.1\textwidth]{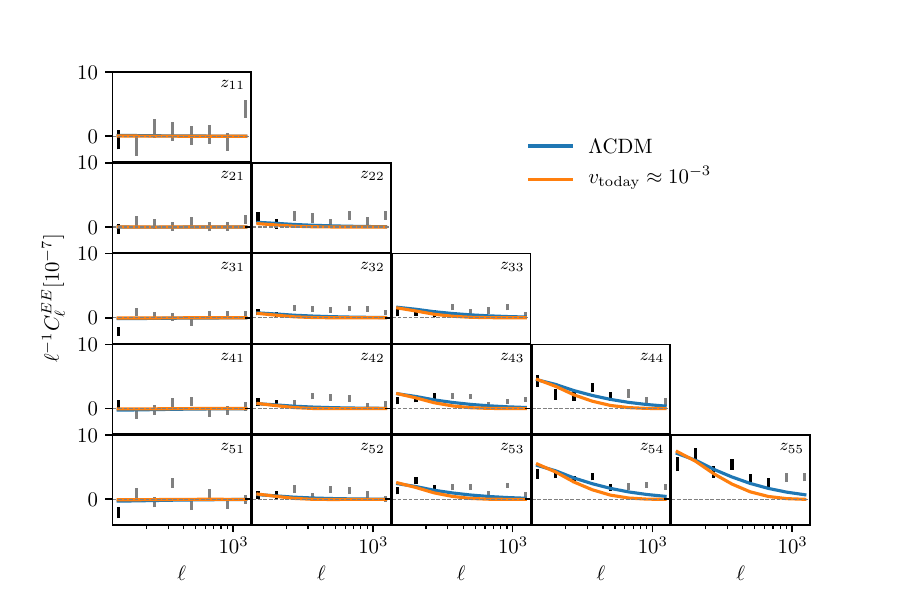}
	\caption{
        KiDS-1000 data overlayed with $C_\ell^{EE}$ best-fit power spectrum calculated for $\Lambda$CDM and from our model for a fixed value of $\vtoday \approx 10^{-3}$. The grey data points have been masked in the analysis since they lie in the non-linear regime. See \S\ref{sec:weakLensing} for more details.}
	\label{fig:shearCls}
\end{figure}

Constraints on $\vtoday$ are easily understood by considering Eq.~\eqref{eq:convergenceCls}, which clearly shows that a suppression in the amount of structure (i.e. $P_m(k, z)$) leads to a suppression in the convergence power spectrum. We show an example of the convergence power spectrum in the $\Lambda$CDM limit ($\vtoday \rightarrow 0$) and the benchmark $\vtoday \approx 10^{-3}$ in Figure~\ref{fig:shearCls}. Note that $C_\ell$'s measuring correlations between higher redshift bins deviate more from $\Lambda$CDM. This is due to the fact that the lensing power is a cumulative effect given by a line-of-sight integral (see Eq.~\eqref{eq:convergenceCls}). Light from the higher redshift tomographic bins travels a longer distance and this gives a bigger difference in the cumulative lensing effect between our model and $\Lambda$CDM.

\section{Model Parameters Constraints}
\label{sec:ModelParameters}

In this section we set the cosmological constraint on the velocity of dark gravitons today in the context of our model and other experimental bounds. In particular, we will discuss what range of model parameters are compatible with current experimental constraints. 

Let us begin by summarizing the constraints we have. First, we have the bound on the velocity today derived in this paper:
\begin{align}
    \label{eq:ModelParamsVelocityConstraint}
    \vtoday = \sqrt{\frac{\delta \mKK}{\mDM}} < 2.2 \times 10^{-4} \qquad \text{(this work, linear cosmology).}
\end{align}

Next, we have the astrophysical bound derived in~\cite{Law-Smith:2023czn}
\begin{align}
    \label{eq:ModelParamsAstroConstraint}
    \mDM < \left(\frac{0.1}{\lambda}\right)^{2/3} \times 100\;\mathrm{keV} \qquad \text{(astrophysics).}
\end{align}

Finally, we also have the bound on the lightest particle in the KK tower from fifth force experiments~\cite{Tan:2016vwu,Lee:2020zjt}:
\begin{align}
    \label{eq:ModelParamsFifthForceConstraint}
    m_1 \equiv \gamma \mKK > 6.6\;\mathrm{meV} \qquad \text{(fifth force experiments).}
\end{align}
In the last inequality, we included an $\mathcal{O}(1)$ number $\gamma$ that is the ratio of the mass of the lightest KK particle to $\mKK$, which is the spacing of the tower at asymptotically large KK numbers. 

Before discussing the constraints these imply on our model parameters, we briefly review the parameters themselves and their physical meaning:
\begin{itemize}
    \item $\lambda$: this is a measure of a suitably average wavefunction overlap between SM fields that live on a (3+1)-brane and KK gravitons that live in an effectively 5D spacetime. This parameter sets the strength of the coupling between the dark gravitons and the SM. As stated earlier, this coupling is of gravitational strength and is determined parametrically by $\MPl$, but the wavefunction overlap changes the value of the coupling by an $\mathcal{O}(1)$ coefficient that we call $\lambda$.
    \item $\beta$: this is a measure of the amplitude of inhomogeneities in the extra dimension. This parameter sets the strength of the coupling between particles in the dark graviton tower. Intra-tower decays are faster for larger $\beta$. 
    \item $\delta$: this describes the wavenumber of the violation of KK momentum in dark graviton decays (as discussed in \S\ref{sec:KKgravitonPheno}).
    \item $\mKK$: this is the spacing between KK particles at large KK quantum numbers and, physically, is the inverse of the size of the extra dimension.
    \item $\gamma$: this is the ratio of the mass of the lightest particle in the tower to $\mKK$, the asymptotic mass spacing.
\end{itemize}
We note that $\mDM$ is not an independent parameter in our model as it can be derived from the remaining parameters. Generally, the parametrics of $\mDM$ are set by equating the decay rate to the Hubble rate (as shown in Eq.~\eqref{eq:DMmass} and discussed in \S\ref{sec:KKgravitonPheno}). For convenience, we repeat that equation below:
\begin{equation}
    \mDM(t) \approx \frac12 \left(\frac{\MPl^4 \mKK}{\delta^3 \beta^4}\right)^{1/7} \frac{1}{t^{2/7}}. \nonumber\\
\end{equation}
We can get a lower bound on $\mDM$ today using Eqs.~\eqref{eq:ModelParamsVelocityConstraint} and~\eqref{eq:ModelParamsFifthForceConstraint}:
\begin{align}
    \label{eq:ModelParamsDMmassBound}
    \mDM > \frac{\delta}{\gamma} \times 100\;\mathrm{keV}.
\end{align}

In summary, our model has five independent parameters $(\lambda, \beta, \delta, \mKK, \gamma)$ and two derived parameters $(\vtoday, \mDM)$ that are determined in terms of the free parameters. In addition, we have three constraints given by Eqs.~\eqref{eq:ModelParamsVelocityConstraint}-\eqref{eq:ModelParamsFifthForceConstraint}. These rule out a portion of parameter space and the goal is to determine the largest region that is compatible with all constraints.

The strongest inequalities we can derive involve weighted products of the parameters $\beta$ and $\delta$. Given values for the remaining parameters, these inequalities will lead to constraints in the $(\beta, \delta)$-plane. The first such inequality is easy to find by rearranging Eq.~\eqref{eq:DMmass} and using the lower bounds on $\mKK$ from Eq.~\eqref{eq:ModelParamsFifthForceConstraint} and $\mDM^{-1}$ from Eq.~\eqref{eq:ModelParamsAstroConstraint}. This gives:
\begin{align}
    \label{eq:ModelParamsBetaDeltaLowerBound}
    \beta^2 \delta^{3/2} > \gamma^{-1/2}(100 \lambda)^{7/3}.
\end{align}

Next, we can also find an upper bound on a weighted product of $\beta$ and $\delta$. This can be obtained by expressing the DM mass in terms of the velocity (as well as $\delta$ and $\beta$) and using the inequalities from Eqs.~\eqref{eq:ModelParamsVelocityConstraint} and~\eqref{eq:ModelParamsDMmassBound}. This gives:
\begin{align}
    \label{eq:ModelParamsBetaDeltaUpperBound}
    \beta^{2/3} \delta^{5/3} < 4 \gamma
\end{align}
The remaining combination of inequalities is given by Eqs.~\eqref{eq:ModelParamsVelocityConstraint} and~\eqref{eq:ModelParamsAstroConstraint}, but it is easy to check that this leads to a bound on $\mKK$ that is much weaker than the experimental constraint Eq. \eqref{eq:ModelParamsFifthForceConstraint} (at least for $\delta,\lambda \sim \mathcal{O}(1)$). 

These last two inequalities provide the strongest relations between the dimensionless parameters of our model. They can be combined (e.g. by eliminating one of $\beta$ or $\delta$) to obtain:
\begin{align}
\label{eq:deltagammalambdaIneq}
    \delta < 0.2\times\gamma\lambda^{-2/3} \\
\label{eq:betagammalambdaIneq}
    \beta > 800 \times\gamma^{-1}\lambda^{5/3} .
\end{align}
That said, it is more illuminating to work with the bounds on the weighted products themselves (Eqs.~\eqref{eq:ModelParamsBetaDeltaLowerBound} and ~\eqref{eq:ModelParamsBetaDeltaUpperBound}), so we proceed to do that.

\begin{figure}[hbtp!]
\centering
	\includegraphics[width=0.45\textwidth]{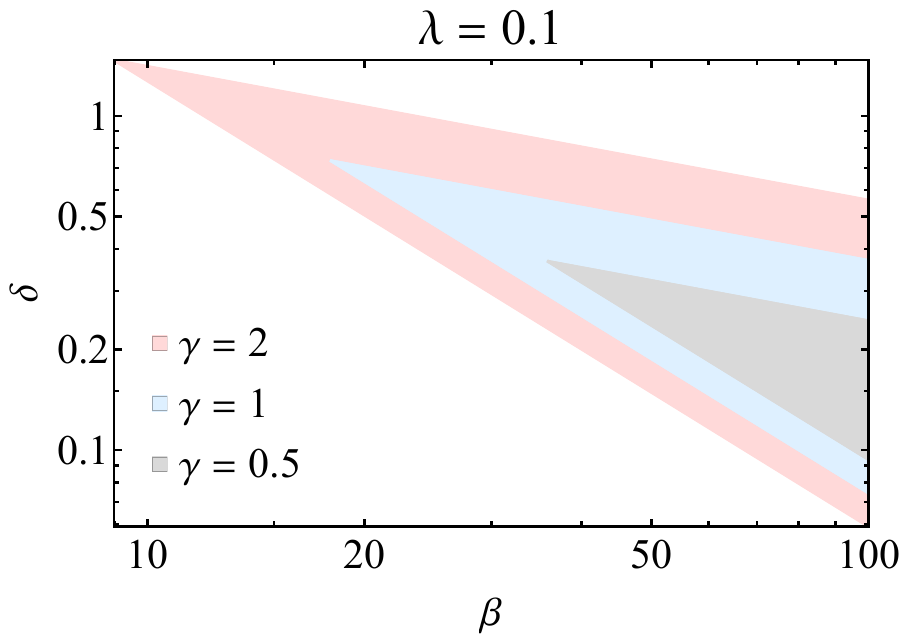}
        \includegraphics[width=0.45\textwidth]{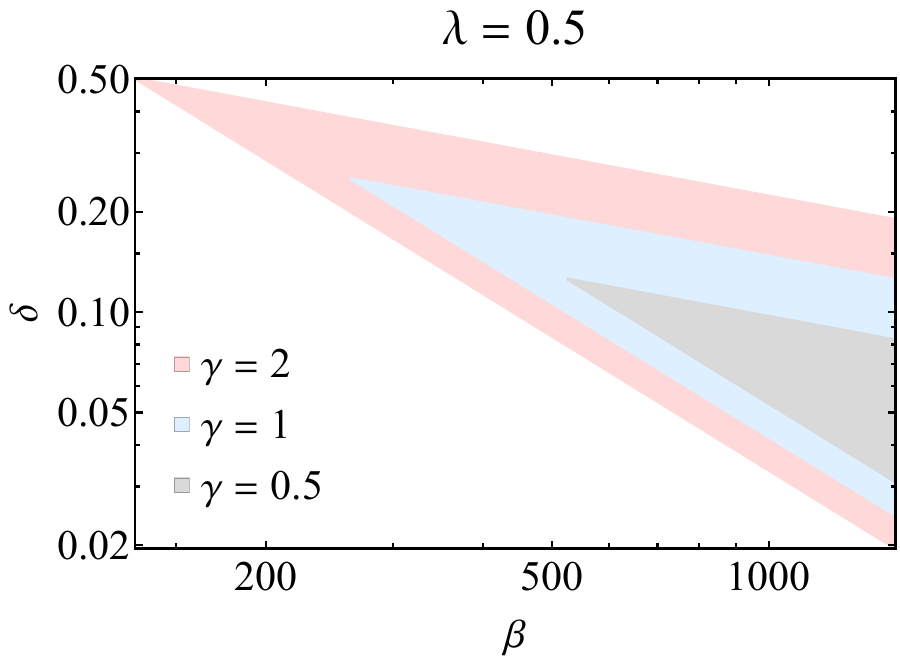}
	\caption{
        The allowed region in the ($\beta$, $\delta$)-plane for various choices of the  parameters $\lambda$ and $\gamma$. Note the different scales on the axes. 
	}
	\label{fig:BetaDeltaRegions}
\end{figure}

To get insight into the allowed values for $\beta$ and $\delta$, we have to choose values the parameters $\lambda$ and $\gamma$. The natural region for the parameter $\lambda$ has been discussed in~\cite{Law-Smith:2023czn}. There, it was found that $\lambda \sim \mathcal{O}(1)$ is allowed by astrophysical observations and large $\lambda$ (which is anyhow unnatural from the point of view of our model) is ruled out. As such, in this work, we take two benchmark values for $\lambda = \{0.1, 0.5\}$. For the parameter $\gamma$, we take three values $\{0.5, 1.0, 2.0\}$. We show the allowed regions in Figure~\ref{fig:BetaDeltaRegions}. 

Finally, using the above constraints we can get a lower bound on the size of the extra dimension (equivalently, an upper bound on $\mKK$). This can be easily obtained using the upper bound on $\vtoday$ from~Eq.~\eqref{eq:ModelParamsVelocityConstraint} and the upper bound on the mass of the dark gravitons from~Eq. \eqref{eq:ModelParamsDMmassBound}. These give:
\begin{align}
    \frac{1}{\gamma} \times 6.6\;\mathrm{meV} <  & \ \mKK < \left(\frac{0.5}{\delta}\right) \left(\frac{0.1}{\lambda}\right)^{2/3} \times 10\;\mathrm{meV} \\
   \gamma \times 30 \mu\mathrm{m} >  & \ l_5 > \left(\frac{\delta}{0.5} \right)\left(\frac{\lambda}{0.1}\right)^{2/3} \times 20 \mu\mathrm{m}.
\end{align}
In particular, note that the overall inequality reduces to that in~\eqref{eq:deltagammalambdaIneq}.

\section{Outlook and Conclusion}
\label{sec:conclusion}

In this work, we studied a model of dark matter motivated by Swampland considerations. In this model, dubbed the `dark dimension', dark matter is composed of a tower of massive spin-2 particles (gravitons) with spacing on the meV scale. Particles in the tower are populated in the early universe and decay to lighter particles as the universe evolves. Each one of these decays gives a kick velocity to the daughter particles, which has an effect on cosmology. It is precisely this effect that we investigated in this work.

We argued why the dark matter kick velocity is time-dependent and used the fluid approximation to study how cosmological perturbations evolve in the presence of this kick velocity. We found, as expected, that the kicks can inhibit the growth of structure on small linear scales and that this leads to a constraint on the velocity today. The constraint originates from effects on the CMB and weak lensing. That said, velocities in the natural range for our model are  viable and may be observed in future experiments. 

For our analysis, we focused on the linear regime. We leave a study of the Dark Dimension model in the non-linear regime for future work. Exploring this question can lead to better understanding of the model and improved constraints. 

Related to the above, large-scale structure surveys will soon present us with a lot more data~\cite{Troja:2022vmg}. It would be interesting to see what that data means in the context of the Dark Dimension model. As emphasized previously~\cite{Montero:2022prj,Gonzalo:2022jac,Law-Smith:2023czn}, the natural parameter range for this model is close to current limits. This view is also reinforced by the analysis in this paper. With new data, the limits will improve and could lead to a detection or ruling out of the model with potential implications for our understanding of quantum gravity.

\subsubsection*{Acknowledgments} 
We would like to thank Tracy Slatyer for valuable discussions.
The work of CV is supported by a grant from the Simons Foundation (602883,CV), the DellaPietra Foundation, and by the NSF grant PHY-2013858. CD is partially supported by the Department of Energy (DOE) Grant No. DE-SC0020223. EG is supported in part by the NSF grants PHY2013988 and PHY1748958. The work of GO work is supported by a Leverhulme Trust International Professorship grant number LIP-202-014. For the purpose of Open Access, the author has applied a CC BY public copyright licence to any Author Accepted Manuscript version arising from this submission.
\begin{appendices} 
\section{Velocity Distribution}
\label{app:velDist}

Let us justify the velocity distribution we use in Eq.~\eqref{eq:distributionFunction}. Since we are interested in the non-relativistic limit, the problem is reduced to the addition of velocity distributions taking into account the vector nature of the particle velocities. Moreover, we are simplifying our problem by assuming that the kick velocity depends only on time. In this appendix we will use another simplification, which is to treat the velocity kicks as happening at discrete time intervals. Each of these time intervals will correspond to a Hubble time where roughly all our dark graviton particles that we had originally have decayed. 

To start with, let us ignore the expansion of the universe and find the evolution of the velocity distribution in flat space ignoring gravity. Let us assume that we start with a collection of particles all of which are at rest so that the initial velocity distribution is given by:
\begin{align}
    \chi(v^2, t_0) \equiv \left.\frac{1}{\rho}\frac{d\rho}{dv^2}\right|_{t=t_0} = \delta(\vec{\bf v}\cdot \vec{\bf v}),
\end{align}
where we emphasize that all distributions we will be considering are isotropic and depend only on the magnitude of the velocity vector $v = |\vec{\bf v}|$. In fact, we will consider the distributions as functions of $v^2$ rather than $v$, which we will see is a more convenient choice. Finally, we point out that the definition of $\chi$ is normalized such that $\int dv^2 \chi(v^2) = 1$. 

At the first time step (the analogue of Hubble in the cosmological scenario), all these particles decay and the daughter particles all get a non-zero velocity. However, given that the magnitude of the kick velocity depends only on time, all the daughter particles will have velocities of the same magnitude but different directions. The velocity distribution then becomes:
\begin{align}
    \chi(v^2, t_1) = \delta(v^2 - v_1^2),
\end{align}
where $v_1$ is the kick velocity obtained in going from $t_0$ to $t_1$. At the first time step, we see that the particles are no longer at rest. Instead, their velocity distribution is given by a $\delta$-function.

At the next time step, in going from $t_1$ to $t_2$, again all particles decay and this imparts a kick velocity $\vec{\bf v}_2$ to be added to the original velocity of the parent particle $\vec{\bf v}_1$. The magnitude of the resulting vector $\vec{\bf v}_3$ is given by:
\begin{align}
    \label{eq:cosineRule}
    v_3^2 = v_1^2 + v_2^2 - 2 v_1 v_2 \cos \theta_{12},
\end{align}
where $\theta_{12}$ is the angle between $\vec{\bf v}_1$ and $\vec{\bf v}_2$. Since the decay is isotropic, the distribution of the cosine term is uniform:
\begin{align}
    \cos \theta_{12} \sim \mathrm{Uniform}(-1,1),
\end{align}
so that $v_3^2$ is also distributed uniformly between $|v_1 - v_2|^2$ and $|v_1 + v_2|^2$:
\begin{eqnarray}
    \chi(v^2, t_3) &=& \mathrm{Uniform}(|v_1 - v_2|^2, |v_1 + v_2|^2) \nonumber\\
    &=& \frac{1}{|v_1 + v_2|^2 - |v_1 - v_2|^2}\left[\Theta(v^2 - |v_1 - v_2|^2) - \Theta(v^2 - |v_1 + v_2|^2)\right] \nonumber\\
    &=& \frac{1}{4v_1 v_2}\left[\Theta(v^2 - |v_1 - v_2|^2) - \Theta(v^2 - |v_1 + v_2|^2)\right],
    \label{eq:uniformChiDistribution}
\end{eqnarray}
where $\Theta$ is the Heaviside function that vanishes for a negative value of its argument and is unity otherwise. The fact that the expression~\eqref{eq:cosineRule} gives $v^2$ is the motivation for taking $\chi$ to be a function of $v^2$.

We see that the distribution becomes wider (going from a $\delta$-function to Uniform) as we step forward in time, and this is expected. Our interest however is in the late-time behaviour of this distribution. So we have to iterate the above procedure for many time steps and describe the late-time result. We will do this numerically but, in order to do that, we will a need a formula to obtain $\chi(v^2, t+\Delta t)$ from $\chi(v^2, t)$ for a general function $\chi$. We do this by relying on this last observation which says that a $\delta$-function in velocity becomes a uniform distribution at the next time step. All we have to do is decompose any distribution of interest into a 'sum' over $\delta$-functions and replace each of these by a uniform distribution.

We now take a general velocity distribution $\chi(v^2, t)$ and deduce the time-evolved distribution $\chi(v^2, t + \Delta t)$ by summing over contributions from all points on $\chi(v^2, t)$ each of which with an amplitude like that given in Eq.~\eqref{eq:uniformChiDistribution}. Let us call the kick velocity $v_k$. We then immediately have:
\begin{align}
    \chi(v^2, t+\Delta t) 
    &= \int_{|v-v_k|^2}^{|v+v_k|^2} d\overline{v}^2 \frac{\chi(\overline{v}^2,t)}{4 \sqrt{\overline{v}^2 v_k^2}}.
\end{align}
As a quick check, we can test the above formula by taking $\chi(\overline{v}^2, t) = \delta(\overline{v}^2 - v_*^2)$, and we recover Eq.~\eqref{eq:uniformChiDistribution}. Using the above formula, and taking $v_k$ to increase as a function of time, it is now straightforward to show (numerically, as we do) that $\chi(v^2, t)$ has a peak around $v_k$ and a width of order $v_k$ as well. This is the late time behaviour of the distribution $\chi(v^2,t)$ that we are after.

We can now use the above equation to get an intuition for how the velocity distribution changes in our model. However, before doing that, we need to include the cosmological expansion which also causes a redshift of the velocity distribution between time steps.

In order to include the cosmological expansion, first observe that the effect of redshift is to transform a distribution $\chi$ in the following way:
\begin{align}
    \chi(v^2, t_0) \rightarrow \left(\frac{a(t)}{a_0}\right)^2 \chi\left(\left(\frac{a(t)}{a_0}\right)^2 v^2, t\right),
\end{align}
where $a_0 = a(t_0)$. The overall factor is important to preserve the normalization $\int \chi dv^2 = 1$ at all times. We can then see that if we were to work with the quantity $w = a(t) v$ and consider distributions of $w^2$ rather than $v^2$, then the problem reduces to the one we already solved in the non-expanding spacetime. Given such a normalized distribution $\eta(w^2, t)$ and the function $a(t)$, we can deduce the corresponding distribution of $v^2$ using:
\begin{align}
\label{eq:chifrometa}
    \chi(v^2, t) = \frac{dw^2}{dv^2}\eta(w^2, t) = a(t)^2\eta(a(t)^2 v^2, t).
\end{align}
In particular the two distributions agree, as they must, when $a(t) \equiv 1$, but the time evolution of $\eta$ under cosmic expansion is trivial, whereas $\chi$ does evolve according to the previous formula. 

To recep, we work with a distribution over $w^2 = a(t)^2 v^2$ that we call $\eta$ and we can deduce the time evolution of $\eta$ using:
\begin{align}
    \eta(w^2, t+\Delta t) 
    &= \int_{|w-w_k|^2}^{|w+w_k|^2} d\overline{w}^2 \frac{\eta(\overline{w}^2,t)}{4 \sqrt{\overline{w}^2 w_k^2}},
\end{align}
where $w_k = a(t) v_k$ is the kick in $w$. Then, given $\eta(w^2, t)$ and the function $a(t)$, we can find the velocity distribution using Eq.~\eqref{eq:chifrometa}.

\begin{figure}[hbtp!]
\centering
	\includegraphics[width=0.6\textwidth]{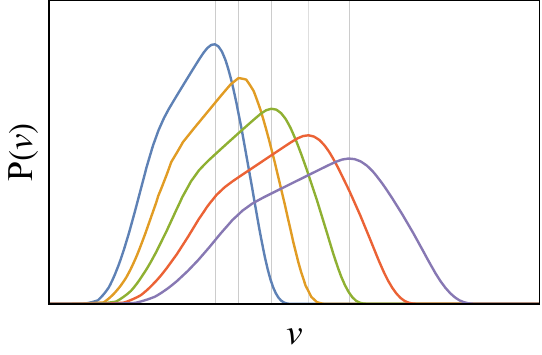}
	\caption{
        The late-time velocity distribution $P(v)$ of decaying particles, where each decay imparts a time-dependent kick velocity as in~\eqref{eq:DMvel}. The different curves show the distribution at time-steps separated by a Hubble time. The vertical grey lines show the peak position. As explained in the text, the distribution peak moves to higher velocity as $\sim t^{1/7}$ and the width also increases so that the ratio of width to peak position is about $0.249$.
	}
	\label{fig:VelocityDistribution}
\end{figure}

The argument is now apparent and we briefly state it here. At late-times, the distribution $\eta(w^2, t)$ behaves just like its static spacetime counterpart. That is to say, it develops a peak around $w_k^2$ and has a width that is also comparable to $w_k^2$. The distribution of $v^2$ is then seen to have a peak that moves with time as $v_k^2 \sim t^{2/7}$. The distribution of velocities, which can then be obtained by a simple change of variables, has a peak that moves as $v_k \sim t^{1/7}$. The width of these distributions is always of order the peak position. We measure this width numerically for the velocity distribution and find:
\begin{align}
    \frac{\mathrm{width}}{\text{peak position}} \approx 0.249,
\end{align}
which is the number we use in our model of the distribution function~\eqref{eq:distributionFunction}. An example of this evolution of the distribution peak and width is shown in Figure~\ref{fig:VelocityDistribution}.

\section{Fluid Approximation}
\label{app:fluidEqs}

We describe perturbations of the graviton dark matter using the fluid approximation. This is adequate in our case since we are dealing with non-relativistic dark matter throughout cosmic history. As such, we do not keep track of the full distribution function and its evolution but only discuss its first few moments, which amounts to using the fluid approximation. The relevant moments are defined in the standard way: 
\begin{align*}
\bar{\rho}(\tau) & =\frac{4 \pi}{a^4(\tau)} \int q^2 d q E(q, \tau) f_0(q,\tau), \\
\bar{P}(\tau) & =\frac{4 \pi}{3 a^4(\tau)} \int q^2 d q \frac{q^2}{E(q, \tau)} f_0(q,\tau), \\
\bar{Q}(\tau) & =\frac{4 \pi}{3 a^4(\tau)} \int q^2 d q \frac{q^4}{E(q, \tau)^3} f_0(q,\tau), \\
\delta \rho(k, \tau) & =\frac{4 \pi}{a^4(\tau)} \int q^2 d q E(q, \tau) f_0(q,\tau) \Delta_0(k, q, \tau), \\
\delta P(k, \tau) & =\frac{4 \pi}{3 a^4(\tau)} \int q^2 d q \frac{q^2}{E(q, \tau)} f_0(q,\tau) \Delta_0(k, q, \tau) \\
(\bar{\rho}+\bar{P}) \theta(k, \tau) & =\frac{4 \pi k}{a^4(\tau)} \int q^2 d q q f_0(q,\tau) \Delta_1(k, q, \tau), \\
(\bar{\rho}+\bar{P}) \sigma(k, \tau) & =\frac{8 \pi}{3 a^4(\tau)} \int q^2 d q \frac{q^2}{E(q, \tau)} f_0(q,\tau) \Delta_2(k, q, \tau) \\
(\bar{\rho}+\bar{P}) \Theta(k, \tau) & =\frac{4 \pi k}{a^4(\tau)} \int q^2 d q q\left(\frac{q}{E(q, \tau)}\right)^2  f_0(q,\tau) \Delta_1(k, q, \tau), \\
(\bar{\rho}+\bar{P}) \Sigma(k, \tau) & =\frac{8 \pi}{3 a^4(\tau)} \int q^2 d q \frac{q^2}{E(q, \tau)}\left(\frac{q}{E(q, \tau)}\right)^2  f_0(q,\tau) \Delta_2(k, q, \tau),
\end{align*}
where $q$ is the particle comoving momentum and $E(q,\tau) \equiv \sqrt{q^2 + a(\tau)^2 m^2}$ is its comoving energy. These definitions relate the multipoles of $\Delta$ to the variables of the fluid description. Using these we can derive the equations governing the fluid perturbations.

We can then proceed as usual by starting with the Boltzmann equation for $d f_0/d t$ and perturbing it by taking $f_0 \rightarrow f_0(1+\Delta)$.  Taking moments, we can then find the perturbation equations, and it is easy to check that these remain unchanged. They give:
\begin{align}
\label{eq:perturbations1}
\dot{\delta} & =-3 \mathcal{H}\left(\frac{\delta P}{\delta \rho}-\omega\right) \delta-(1+\omega)[\theta-3 K_0] \\
\label{eq:perturbations2}
\dot{\theta} & =-\frac{\dot{\omega}}{1+\omega} \theta+\mathcal{H}(-1+3 \omega) \theta +\frac{\delta P / \delta \rho}{1+\omega} k^2 \delta-k^2 \sigma+ 3k K_1 \\
\label{eq:perturbations3}
\dot{\sigma} & =-\frac{\dot{\omega}}{1+w} \sigma+\mathcal{H}(-2+3 w) \sigma +\mathcal{H} \Sigma+\frac{4}{15} \Theta + X - \frac{2}{1+w} K_2 \left(-5w + \frac{\bar{Q}}{\bar{\rho}}\right).
\end{align}
In the last equation, we have included a term called $X$ that encapsulates the effect of truncation (i.e. $\Delta_3 = 0$) versus using the Ma-Bertschinger~\cite{Ma:1995ey} approximation:
\begin{align*}
  \Delta_3 \approx \frac{5 E}{q k \tau} \Delta_2-\left(\Delta_1-\frac{1}{3} \alpha k \frac{E}{q} \frac{d \ln f_0}{d \ln q}\right).
\end{align*}
In the case where a truncation is used, we have $X = 0$, otherwise:
\begin{align*}
  X = -\frac{3}{\tau}\sigma + \frac25 \Theta -\frac{2}{5(1+w)}\alpha k^2 \left(-5 w + \frac{\bar{Q}}{\bar{\rho}}\right).
\end{align*}
In addition, $\alpha = 0$ in the Newtonian gauge and $\alpha = (\dot{h}+6 \dot{\eta}) /\left(2 k^2\right)$ in the synchronous gauge. Finally the terms $K_i$ depend on whether we are using the Newtonian or synchronous gauge and they are given by:
\begin{align}
K_0 &= \left\{
\begin{array}{lc}
    \dot{\phi} & \qquad \text{Newtonian} \\
    \frac16 \dot{h} &\qquad \text{synchronous}
\end{array}
\right. \\
K_1 &= \left\{
\begin{array}{lc}
    \frac13 k\psi & \qquad \text{Newtonian} \\
    0 &\qquad \text{synchronous}
\end{array}
\right. \\
K_2 &= \left\{
\begin{array}{lc}
    0 & \qquad \text{Newtonian} \\
    \frac{1}{15}(\dot{h} + 6\dot{\eta}) \dot{h} &\qquad \text{synchronous},
\end{array}
\right. 
\end{align}
where the metric perturbations $(h, \eta, \psi, \phi)$ are defined in the standard way (e.g. as in~\cite{Ma:1995ey}). The system of equations in Eqs.~\eqref{eq:perturbations1}-\eqref{eq:perturbations3} is not closed, and we close it in the same way that is discussed in~\cite{Lesgourgues:2011rh}.

We work mostly in the synchronous gauge. This is an incomplete gauge fixing and we are allowed to perform gauge transformations that are time-independent. This can be a nuisance because once we get a solution, we have to check whether it can be removed by a gauge-transformation. We usually resolve this ambiguity by demanding that the DM velocity perturbation $\theta = 0$. This can be confusing because all our effects come from $\theta \neq 0$. How are we then to understand the use of synchronous gauge in this case?

First, let us review why we can set $\theta = 0$ in the case of CDM. The equation of motion for $\theta$ implies that $\dot{\theta} = 0$ for CDM (i.e. with $w = 0$ and $\delta P = 0$ and when anisotropic stresses vanish). We then have that $\theta(\vec{x})$ depends on $\vec{x}$ only and we can use the remaining gauge-freedom to set $\theta = 0$. This completes the gauge-fixing when we have a CDM component.

In our case, we assume the presence of a very subdominant CDM component, with $\Omega_{c} \sim 10^{-10}$ (this is in fact already implemented in CLASS). The gauge freedom is then removed by removing the velocity perturbation of this component. Because of its extremely low energy density, this fictitious component does not affect the evolution of the universe in any way that has bearing on experimental constraints. It is then clear that the velocity perturbation of our dark gravitons is physical.

This leaves open one question: even in the presence of CDM, why not assume there is a subdominant (also cold) component and then treat the CDM velocity perturbation as physical? The answer is that, as reviewed before, for CDM we have the equation $\dot\theta = 0$ where we can afterwards set $\theta = 0$ by a gauge transformation. So whenever we get a non-zero constant-in-time $\theta$ for CDM, we can remove it by a gauge transformation. (Typically there should be one linear combination of the two velocity perturbations that can be removed and another one that is physical). What remains then is the velocity perturbation of the subdominant component which again has little bearing on cosmological evolution. In the case of dark gravitons, the $\theta$ perturbation is not constant in time and it cannot be removed by a time-independent gauge transformation. The assumption of the existence of an additional subdominant component with a vanishing velocity perturbation completely fixes the synchronous gauge. 

Finally, we quote the expression for the constant pressure, pseudo-pressure and energy density used in Eqs.~\eqref{eq:wfld} and~\eqref{eq:soundspeedfld}. These can be obtained by integrating the distribution function in Eq.~\eqref{eq:distributionFunction} as in the expressions for $\overline{P}, \overline{\rho}$ and $\overline{Q}$ in the non-relativistic limit. The result then gives $p_0, \rho_0$ and $\mathfrak{p}_0$ after factoring out the appropriate powers of $a(t)$ and $\vtoday (t/t_{\rm age})^{1/7}$. These are:
\begin{align}
    p_0 &= \frac{2 \pi \sqrt{\eta} m^4}{3} \left(
    2 e^{-1/2\eta} \sqrt{\eta}(1+5\eta) +\sqrt{2\pi}(1 + 6\eta + 3\eta^2)(1 + \mathrm{Erf}(1/\sqrt{2\eta}))
    \right)\\
    \rho_0 &= 2\pi \sqrt{\eta} m^4 \left(
    2 e^{-1/2\eta} \sqrt{\eta} + \sqrt{2\pi}(1+\eta) (1 + \mathrm{Erf}(1/\sqrt{2\eta}))
    \right)\\
    \mathfrak{p}_0 &= \frac{2 \pi \sqrt{\eta} m^4}{3} \left(
    2 e^{-1/2\eta} \sqrt{\eta}(1+3\eta)(1+11\eta) + \sqrt{2\pi}(1 + 15\eta + 15\eta^2(3 + \eta))(1 + \mathrm{Erf}(1/\sqrt{2\eta}))
    \right)
\end{align}
where we take $\eta = 0.25$ as per the discussion in Appendix~\ref{app:velDist}. Here $m$ is the DM mass, but it drops out of all expressions in the non-relativistic limit of interest to us (see for example Eqs.~\eqref{eq:wfld},~\eqref{eq:soundspeedfld} and~\eqref{eq:distributionFunction}).

\end{appendices}

\bibliographystyle{unsrt}
\bibliography{main}

\end{document}